\documentclass{llncs}
\usepackage{amsfonts}       %
\usepackage{amsmath}
\usepackage{graphicx}
\usepackage{subcaption}
\usepackage{tikz}
\usepackage{float}
\usepackage{booktabs}       %

\newcommand{\R}{\mathbb{R}}

\usepackage{hyperref}
\usepackage{url}

\title{Latent-space disentanglement with untrained generator networks for the isolation of different motion types in video data}

\author{Abudllah\inst{1} \and
Martin Holler\inst{2} \and Karl Kunisch\inst{2} \and Malena Sabate Landman\inst{3}}

\institute{Department of Mathematics, The Chinese University of Hong Kong \email{abdullahuom1@yahoo.co} \and 
Institute of Mathematics and Scientific Computing, University of Graz \email{\{martin.holler,karl.kunisch\}@uni-graz.de}
\and Department of Mathematics, Emory University \email{malena.sabate.landman@emory.edu}}

\usepackage{fancyhdr}
\begin{document}

\maketitle

\textbf{This preprint has not undergone peer review or any post submission improvements or corrections. The Version of Record of this contribution is published in Lecture Notes in Computer Science, and is available online at 
\href{https://doi.org/10.1007/978-3-031-31975-4\_25}{https://doi.org/10.1007/978-3-031-31975-4\_25}}

\begin{abstract}
Isolating different types of motion in video data is a highly relevant problem in video analysis. Applications can be found, for example, in dynamic medical or biological imaging, where the analysis and further processing of the dynamics of interest is often complicated by additional, unwanted dynamics, such as motion of the measurement subject.
In this work, it is empirically shown that a representation of video data via untrained generator networks, together with a specific technique for latent space disentanglement that uses minimal, one-dimensional information on some of the underlying dynamics, allows to efficiently isolate different, highly non-linear motion types. In particular, such a representation allows to freeze any selection of motion types, and to obtain accurate independent representations of other dynamics of interest.
Obtaining such a representation does not require any pre-training on a training data set, i.e., all parameters of the generator network are learned directly from a single video.

\end{abstract}

\section{Introduction}
Processing motion information in a time series of images is a classical but still very active research topic in computer vision and computational imaging, with a plethora of applications ranging from autonomous driving to biological and medical imaging. 
In this context, one can separate three (strongly interconnected) directions of research: i) Motion reconstruction, which aims at reconstructing dynamic image data from incomplete or indirect measurements, with applications  for instance in dynamic magnetic resonance (MR) imaging or dynamic positron-emission tomography (PET), see \cite{holler17ictgvmri,rahmim2009four} for examples. ii) Motion estimation, which aims to estimate and represent motion between different frames. This is one of the most classical problems in computer vision and often addressed, for instance, via optical flow estimation, see \cite{fortun2015optical} for a review. iii) Motion correction, which aims to correct for motion, typically via registration techniques. The latter are again a well-established but still very active research topic, in particular in the context of medical imaging such as MR imaging, PET, computed tomography (CT), see for instance \cite{oliveira2014medical}.

Naturally, close connections between these three direction of research exist: Having estimated motion fields available is crucial for motion correction, motion correction is strongly interconnected with image reconstruction, and adequately reconstructed time series data is the basis of classical motion estimate techniques. 

In the past years, deep learning based methods have enabled a significant progress in all of the above research direction connected to motion in time series of images, see \cite{bustin2020compressed}, \cite{tu2019survey} and \cite{fu2020deep} for recent review papers on deep learning techniques for reconstruction, motion estimation and motion correction, respectively. 

A task that is related but still different to the above-described research directions is the \emph{isolation} of different types of motion. By this, we mean the following general problem setting: Given a video with different types of motion, synthesize a new video that does not show all types of motion, but only a subset of motion-types that is relevant for further processing. A generic area of applications where this problem is relevant is medical imaging, e.g., MR imaging or PET. Here, a concrete example is a video showing cardiac motion together with additional motion resulting from breathing or patient movement, and the goal is to obtain a video showing only the isolated cardiac motion for further analysis. 

While the task of isolating motion is strongly related to image registration techniques, the latter cannot directly be applied here since, even in the case where motion fields that register each frame of the time series to a representative template are available, it is still a highly non-trivial problem to decompose such motion fields into different components corresponding to different types of motion. In the specific context of measuring myocardial perfusion, there exist approaches that aim to overcome this by combining registration techniques, e.g., with independent- or principle component analysis \cite{WOLLNY20121015,Adkinson14_robust_pca_motion_correction} or sparsity priors on the perfusion dynamics \cite{Mathews15_compressed_sensing_motion_correction}. Such approaches, however, require a specific structure of the dynamics of interest (i.e. perfusion) and an explicit modeling of deformations. The extension of such techniques, e.g. to allow for the isolation of two types of morphological motion such as cardiac and respiratory motion, is non-trivial.

A more generally applicable alternative to such approaches would be to use a non-linear extensions of ICA or variational autoencodes, see for instance \cite{hyvarinen1999nonlinear,khemakhem2020variational} or \cite{2021disentangling} and the references therein for more structured models. To the best knowledge of the authors, however, it is not clear to what extend non-linear ICA can be applied for isolating motion types based on a single video, since these approaches are typically applied in different contexts and for large datasets.

In this work, we present a new idea for motion isolation that we believe to have great potential for diverse applications,
since it is generally applicable to different medical imaging modalities. Moreover, it can be directly incorporated in (variational) image reconstruction methods and, contrary to classical registration techniques, it does not rely on explicit modeling of different motion types. 
The main idea behind our approach is to employ untrained generator networks, together with a specific technique for latent space disentanglement, for motion isolation. More specifically, we consider the optimization of a generator network to represent a given time series of images, where different latent space variables are forced to independently explain the different types of motion. The latter is achieved by incorporating one-dimensional information on all but one of the different motion types present in the video.

The goal of this paper is to outline this new idea of motion isolation, and present a first numerical evaluation based on synthetic examples and semi-synthetic examples with real dynamic cardiac MR image data, where we isolate cardiac motion from respiratory motion.
The one dimensional information on some of the underlying motion types in this case corresponds to a scalar describing the cardiac- or breathing state, a signal which in practice can easily be obtained for instance by simultaneous electrocardiogram (ECG) or chest-displacement measurements.

The use of untrained generator networks for image representation was popularized by \cite{ulyanov2020dip}, which also partially inspired our work. Since the appearance of \cite{ulyanov2020dip}, many works employing the deep image prior for image reconstruction in various applications appeared. %
More recently, also works that employ the deep image prior for representation and reconstruction of dynamic MR data appeared, see for instance \cite{hyder2020generative,yoo2021time}. Existing works, however, focus on reconstruction and do not employ a specific latent-space disentanglement as proposed here, which is the main ingredient for not only representing but also isolating different motion types.

Latent space disentanglement is in turn an active research topic in the context of GANs, see for instance \cite{chen2016infogan} for a seminal work on using latent space disentanglement to learn interpretable representations, and, e.g., \cite{tulyaov2018mocogan} for a work on decomposing motion and content in videos. But again, also in the context of GANs, to the best knowledge of the authors, a method capable of isolating different motion types in video data based on a single video does not exist.

\section{Method}\label{sec:Method}
We introduce the proposed approach in a general setting (including possibly indirect observations of the image data) first, and then specify its concrete application to isolating respiratory and cardiac motion in dynamic images that were obtained with MR imaging.

Consider a linear, discretized dynamic imaging inverse problem with data $(y_t)_{t=1}^T \subset \R^M$ and linear operators $A_t \in \mathcal{L}(\R^{N_1\times N_2},\R^M)$ of the form 
\begin{equation}
y_t = A_t x_t, \quad {t=1,...,T},
\end{equation}
where the unknown is a sequence of images $(x_t)_{t=1}^T$ with $x_t \in \R^{N_1\times N_2}$ for each $t$. Following the basic idea introduced in \cite{ulyanov2020dip} for static images, we consider each frame $x_t\in \R^{N_1\times N_2}$ to be the output of a generator $x_t= G_{\theta}(z_t)$, whose parameters we want to learn only from the given data. In particular, we consider generator networks $G_\theta : \R^q \rightarrow  \mathbb{R}^{N_1 \times N_2}$ of the form
\begin{equation}\label{eq:generator_network}
G_\theta(z_t) = \theta^1_L \ast \sigma_{L-1}(\theta^1_{L-1}\ast (...\sigma_1(\theta^1_1 \ast z_t + \theta_1^2)...) + \theta_{L-1}^2) + \theta_L^2,
\end{equation}
with time-independent parameters $\theta^{i}_{j} \in \Theta$, for $1 \leq i \leq 2$ and $ 1\leq j \leq L$ and pointwise nonlinearities $\sigma_{1},\ldots,\sigma_{L-1}$. The generator network maps the latent space $\mathbb{R}^{q}$ to the image (frame) space $ \R^{N_1 \times N_2}$. 
Assuming the dynamic image sequence $(x_t)_{t=1}^T$ contains $m \in \{2,3,\ldots\}$ independent types of motion, we split the latent variable $z \in \R^q$ into a time-independent part $z^0 \in \R^{q-m}$ and $m$ time-dependent variables $(z^i_t)_{t=1}^T$ with $z^i_t  \in \R$, $i=1,\ldots,m$. We further assume that all the time dependent variables $(z^i_t)_{t=1}^T$ except $(z^1_t)_{t=1}^T$ are given as $(\hat z^i_t)_{t=1}^T$ from some one-dimensional a-priori information on the state of the respective types of motion (e.g. from electrocardiograms or
chest-displacement measurements). 

We then reconstruct the image sequence $(x_t)_{t=1}^T$ together with the network parameters $\theta$ and the time-dependent variable $(z^1_t)_{t=1}^T$ via solving
\begin{equation}\label{eq:general_opt_problem}
((\hat{z}^1_t)_{t=1}^T,\hat{\theta})  \in  \arg \min_{(z^1_t)_{t=1}^T,\theta} \frac{1}{T}\sum_{t=1}^{T}\|y_t - A_t G_\theta (\hat z^0,z^1_t,\hat z^2_t\ldots,\hat z^m_t)\|^2_2,
\end{equation}
where $\|\cdot \|_2$ is the Euclidean norm (but can be a more general loss) and $\hat z^0\in \R^{q-m}$ is a randomly initialized, fixed static latent variable.
Once a solution is obtained, we do not only obtain the reconstructed images sequence $(\hat{x}_t)_{t=1}^T$ via $\hat{x}_t =  G_{\hat \theta }(\hat z^0,\hat z^1_t,\ldots,  \hat z^m_t)$, but, more importantly, can generate image sequences $(\hat{x}^i_t)_{t=1}^T$ for $i=1,\ldots,m$, which we expect to contain only the $i$th type of motion with all others being fixed, via
\begin{equation} \label{eq:fixing_frames}
 \hat x^i_t = G_{\hat \theta} (\hat z^0,\hat z^1_{h_1},\ldots,\hat z^{i-1}_{h_{i-1}}, \hat z^i_{t}, \hat z^{i+1}_{h_{i+1}},\ldots,\hat z^m_{h_m}),
\end{equation}
where $h_1,\ldots,h_{i-1},h_{i+1},\ldots,h_m$ are fixed reference frames.  

As a concrete example, in this paper we consider the application of this general approach to isolating cardiac motion from respiratory motion in dynamic images obtained from MRI, where one-dimensional information about the respiratory state (e.g. from measurements of the chest displacement) is available. In this case, $m=2$, and the latent variable $z_t \in \R^q$ at time $t$ is decomposed as $z_t = (z^0,z^1_t,z^2_t)$ with $z^2_t $ known (and given as $(\hat z^2_t )_{t=1}^T$). As our focus is on motion isolation rather than reconstruction, we further assume the reconstructed dynamic image sequence to be available, i.e., $A_t$ is the identity, noting that a generalization of our approach to reconstruction does not pose any conceptual difficulties. In summary, this yields the following optimization problem
\begin{equation}\label{eq:specific_opt_problem}
((\hat{z}^1_t)_{t=1}^T,\hat{\theta})  \in  \arg \min_{(z^1_t)_{t=1}^T,\theta} \frac{1}{T}\sum_{t=1}^{T}\|y_t - G_\theta (\hat z^0,z^1_t,\hat z^2_t)\|^2_2.
\end{equation}

\paragraph{Algorithmic strategy.}%

To solve the minimization problem \eqref{eq:specific_opt_problem}, we use Pytorch \cite{paszke2017automatic_diff_pytorch} and the ADAM optimizer %
 with default settings. To achieve a good minimization of the loss, and in particular stability w.r.t varying random initializations, we iteratively reduce the learning rate after a fixed number of epochs, track the network parameters and latent variables that achieve the minimal loss, and export those parameters and variables as the optimal solution (instead of the variables of the last iterate). Note that this does not cause much computational overhead due to the rater small dimensionality of our network (see Section \ref{sec:experiments}).

\paragraph{Baseline method.} In order to have a baseline method for comparison, we also implemented an approach for motion isolation that is based on independent component analysis as follows: Any given video $(x_t)_{t=1}^T$ is first rearranged to a matrix $A \in \R^{(N_1N_2) \times T}$, with the rows representing spatial dimension and columns representing the temporal dimension. This matrix is then decomposed as $A = SV$ with $S  \in \R^{(N_1N_2) \times 2}$ and $V \in \R^{2 \times T}$ using the FastICA implementation of \cite{scikit-learn}, where the two rows of $V$ correspond to the two independent components. Based on manual labeling, one row (say the first one) of $V$ is then assigned to respiratory motion, and the other to cardiac motion. Given this, an approximation of the respiratory motion at any state $t_1$ and of the cardiac motion at any state $t_2$ can then obtained as $(S_{:,1}V_{1,:})_{:,t_1} + (S_{:,2}V_{2,:})_{:,t_2}$ (using standard matrix indexing). Fixing one of these states and letting the other one run from $1$ to $T$, a video of only respiratory or cardiac motion can be obtained.

Note that, while this method is a simple, baseline approach to separate data into different, independent motion components, it has several limitations: Since the separation is linear, it can be expected to (approximately) work only for certain types of motions and images. Furthermore, the method can only be used for fully-sampled videos, but not for indirect measurements as in \eqref{eq:general_opt_problem}. Advantages of the method on the other hand are its simplicity and that it does not require any a-priori information on the motion.

\section{Numerical Experiments} \label{sec:experiments}
In this section the results of different experiments concerning dynamic images with respiratory and cardiac motion are presented to illustrate the behavior of the new method. In particular, we present a synthetic phantom example and four semi-synthetic examples where real dynamic cardiac MR images were enriched with synthetic respiratory motion. Note that a synthetization of one of the motion types is necessary in order to have a ground-truth with isolated motion.

For all experiments, we assume one-dimensional information, henceforth referred to as motion triggers, about the respiratory motion to be given. For the phantom images, the triggers for the respiratory motion are assumed to be given exactly, where for the real MR images we provide the method only with a perturbed version of the original respiratory motion triggers.

To assess the quality of the motion isolation, we compute the relative error norms ($\mathbf{E}_h^1$ and $\mathbf{E}_h^2$) of the dynamic images $\hat x^1 = (\hat x^1_t )_{t=1}^T$, $\hat x^2= (\hat x^2_t )_{t=1}^T$ containing just one reconstructed kind of motion, where
\begin{equation}\label{er}
	 	\mathbf{E}_h^1 = \|\hat{x}^{1}- x^1_{\text{true}}\|_2/\| x^1_{\text{true}}\|_2, \quad \mathbf{E}_h^2 = \|\hat{x}^{2}- x^2_{\text{true}}\|_2 / \|x^2_{\text{true}}\|_2,
\end{equation}
and $x^i_{\text{true}}$ is the ground truth showing only the $i$th type of motion, i.e., cardiac motion for $i=1$ and respiratory motion for $i=2$. Here, the subscript $h$ refers to the frame at which the other motion state is fixed, see \eqref{eq:fixing_frames}. %

In principle, as described in Section \ref{sec:Method}, our method (as well as the ICA-based baseline approach) allows to freeze one kind of motion at any state, and generate images containing only the second kind of motion, as long as a sufficient mixing of motions was observed. In practice, the choice of the freezing frame $h$ has an impact on the performance of the single motion reconstruction (though for the phantom at an overall rather low error regime). %
To allow for a fair comparison, for each method we always show the result for fixing the motion state that provides the best performance with respect to the ground truth.

For all experiments with our as approach shown in the paper, we repeated the experiment 20 times with 20 different random seeds, and show the result whose performance w.r.t. the error measure $\mathbf{E}_h^1 $ is closest to the median performance. 
Quantitative error measures for all experiments are provided in Table \ref{tbl:seed_errors}. Videos showing results for three different experiments are available in the supplementary material of this work.
The source code to reproduce all experiments will be available after acceptance of this paper.
All our experiments were conduced on a workstation with an AMD Ryzen 7 3800X 8-Core Processor and 32 GB of memory, using a Nvidia RTX 3090 GPU with 24 GB of memory. The smallest and largest experiment considered here (solving \eqref{eq:specific_opt_problem} with phantom and real cine MR data, respectively) took around 19 seconds and 3.8 minutes, respectively.

\subsection{Synthetic data}
\begin{figure}[t]
	\centering	
	\includegraphics[width=0.65 in]{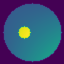} %
	\includegraphics[width=0.65 in]{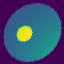} %
	\includegraphics[width=0.65 in]{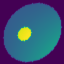} %
	\includegraphics[width=0.65 in]{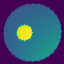} %
	\includegraphics[width=0.65 in]{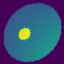} %
	\includegraphics[width=0.65 in]{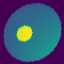} %
	\includegraphics[width=0.65 in]{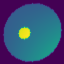} %
	\caption{Selected synthetic data frames displaying different phases of respiratory and cardiac movement.}
	\label{oriph}
\end{figure}

\begin{figure}[t]
	\centering
	\begin{subfigure}[b]{0.49\textwidth}
	\centering
		\begin{tikzpicture}
		\node[inner sep=0pt] (russell) at (0,0)
		{\includegraphics[height= 0.7 in]{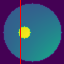}};
		\node[inner sep=0pt] (whitehead) at (2.5,0)
		{\includegraphics[width= 1 in, height= 0.7 in]{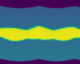}};
		\node[node distance=0cm, xshift=2.5cm, yshift=-1.2cm,font=\color{black}] {{\tiny time}};
		\draw[->] (1.4,-1.1) -- (3.6,-1.1);
		\end{tikzpicture}
		\label{fig:simpvid}
	\end{subfigure}
	\begin{subfigure}[b]{0.49\textwidth}
	\centering
		\includegraphics[height=0.8 in, width=\textwidth, trim = 2.5cm 1cm 3cm 0, clip]{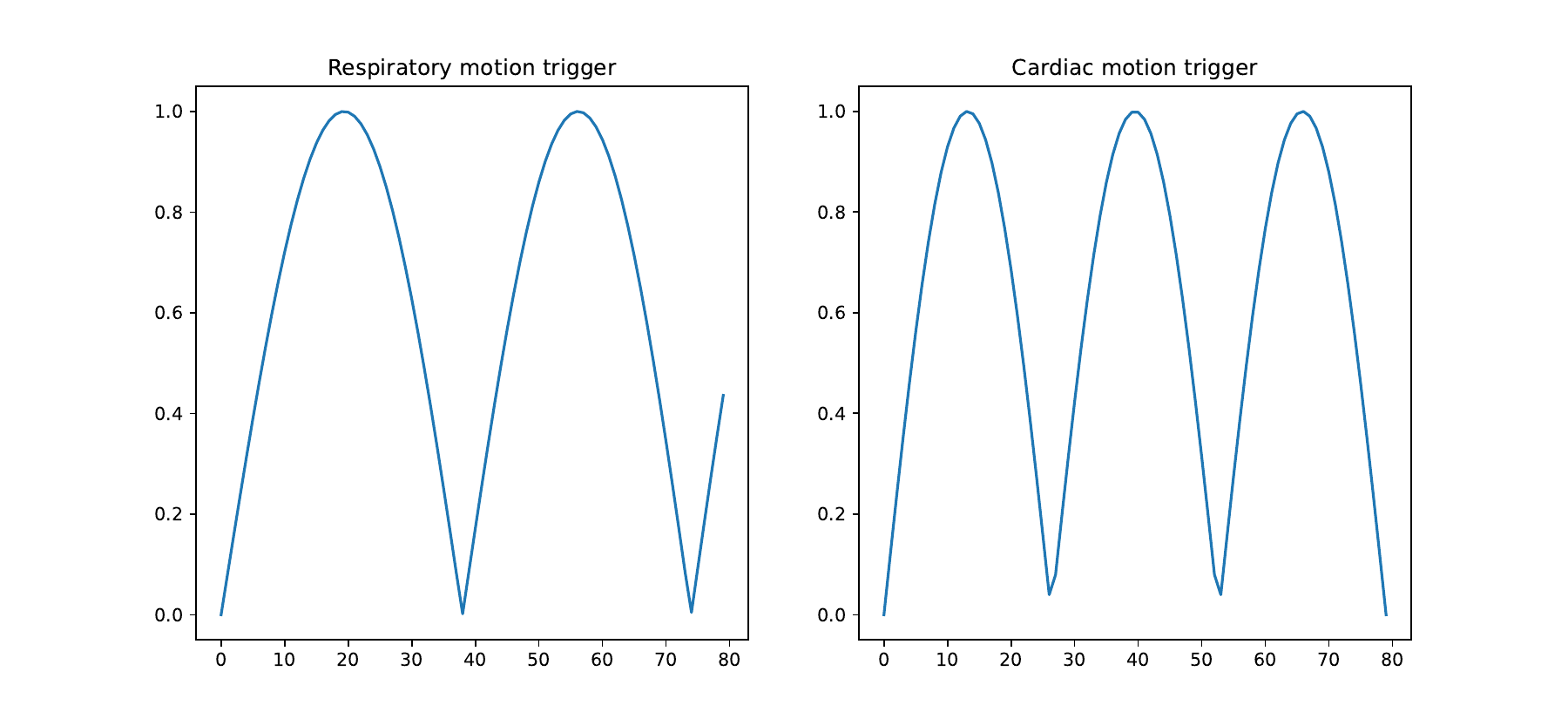}
		\label{origcount}
	\end{subfigure}
	\caption{Representation of the time evolution for the synthetic data. Left: Compact video representation. Right: Ground truth motion triggers.\label{fig:phantom_ground_truth}}
\end{figure}

The first test problem, consisting of 80 frames with  64$\times$64 pixels, corresponds to a synthetic example displaying two nested disks and is represented in Figure \ref{oriph}. A more compact representation of this dynamic image can be observed in Figure \ref{fig:phantom_ground_truth} (left), where a vertical slice of the image (marked in red on the left image), is displayed over time (and can be seen on the right image) clearly showing temporal changes. This representation will be used throughout the paper.

In this example, three cardiac motion cycles are simulated by dilation of the internal disk while approximately two simulated breathing motion cycles are represented by shearing of the whole image. Note that the size of the frames over time is maintained constant. The ground truth one-dimensional motion information, i.e., the motion triggers, that was used to parametrize the different types of motions is displayed in Figure \ref{fig:phantom_ground_truth} (right). Note that the motion trigger for respiratory motion is provided to our method via the latent variable $(z^2(t))_{t=1}^T$, while the motion trigger for cardiac motion is an unknown optimization variable.

The generator modeling the solution for this example, as defined in Equation \eqref{eq:generator_network}, corresponds to a standard deep convolutional neural network (CNN) with 5 layers, where transpose two-dimensional convolutions are used for all convolutions in \eqref{eq:generator_network} and no biases are used. The network parameters are given as follows. Number of channels: $[64,128,64,32,16,1]$, (square) kernel size: $[4,4,4,4,4]$, stride: $[1,2,2,2,2]$, padding: $[0,1,1,1,1]$, activation functions: $[\text{Tanh},$ $\text{LeakyReLU},$ $\text{Tanh},$ $\text{LeakyReLU},$ $\text{Tanh}]$. In total, the network has $3.03360 \times 10^{5}$ parameters.
The latent space $Z \in \mathbb{R}^{64}$ is split into 62 static components and 2 dynamic components. Blocks of $[4000,4000,4000,4000]$ epochs with learning rates $[0.01,0.005,0.001,0.0005]$ are used for the Adam optimizer. 
The latent variables are initialized randomly from a uniform distribution on $[0, 1)$, and the network weights are initialized with Pytorch self-initialization.

\begin{figure}[t]
	\begin{subfigure}[b]{0.48\textwidth}
		\centering
\includegraphics[height=.7 in]{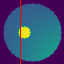}
\includegraphics[width=1 in, height= 0.7 in]{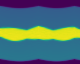}
	\end{subfigure}
	\begin{subfigure}[b]{0.48\textwidth}
		\centering
		\includegraphics[height=0.8 in,trim=1cm 0.6cm 1cm 0, clip]{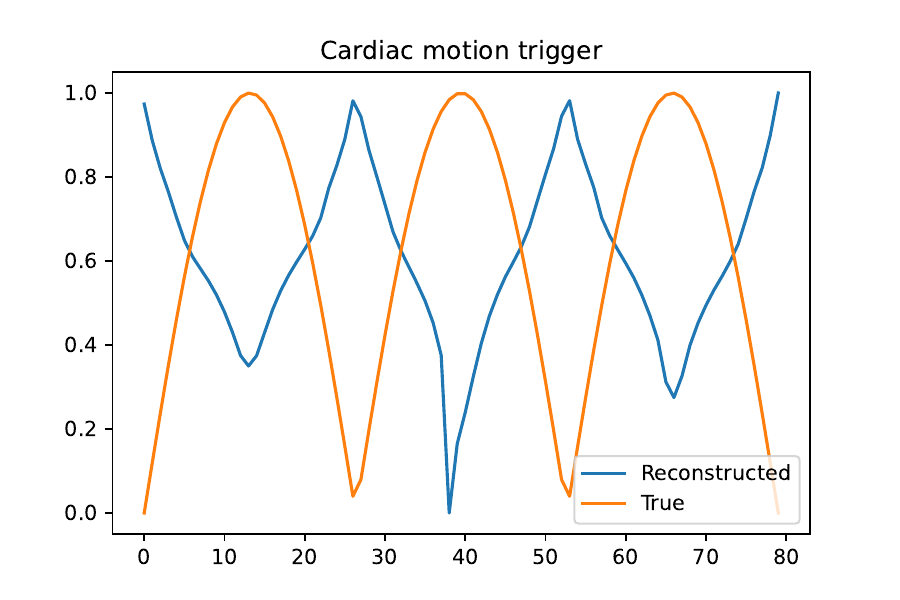}
	\end{subfigure}
	\caption{Synthetic video reconstruction with full motion. Left: Video data, right: heart motion triggers (rescaled).} \label{Ex1_fullmotion}
\end{figure}

The reconstruction of the dynamic image data resulting from the numerical solution of \eqref{eq:specific_opt_problem} is shown in Figure \ref{Ex1_fullmotion}. Even though a representation of the given data with mixed motion is not our primary goal, it can still be observed that resulting video is visually very similar to the ground truth, giving evidence that one-dimensional information on one of the movements is enough to disentangle the latent space without a significant degradation in representing the data. Regarding the reconstructed motion triggers for cardiac motion (the variable $(z^1_t)_{t=1}^T$, see Figure \ref{Ex1_fullmotion} right), we do not expect an accurate reconstruction due to unavoidable ambiguities how such motion can be represented. Nevertheless, one would expect to see a periodic behavior in the reconstructed motion trigger, with the same period as the ground truth. Except for one larger deflection of the motion trigger during one period (which is probably canceled out by the nonlineraties in the network) this can indeed be observed.

\begin{figure}[t]
\centering
		\includegraphics[width=0.19\textwidth, height= 0.7 in]{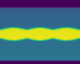}
		\includegraphics[width=0.19\textwidth, height= 0.7 in]{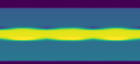}
		\includegraphics[width=0.19\textwidth, height= 0.7 in]{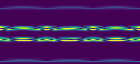}
		\includegraphics[width=0.19\textwidth, height= 0.7 in]{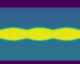}
		\includegraphics[width=0.19\textwidth, height= 0.7 in]{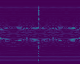}
		\includegraphics[width=0.19\textwidth, height= 0.7 in]{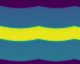}
		\includegraphics[width=0.19\textwidth, height= 0.7 in]{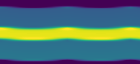}
		\includegraphics[width=0.19\textwidth, height= 0.7 in]{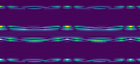}
		\includegraphics[width=0.19\textwidth, height= 0.7 in]{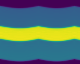}
		\includegraphics[width=0.19\textwidth, height= 0.7 in]{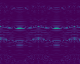}

	\caption{Compact representation of the synthetic video reconstructed with only cardiac (top row) and only respiratory (bottom row) motion. From left to right: Ground truth, ICA-based motion isolation with corresponding difference image, proposed motion isolation technique with corresponding difference image.}
	\label{fig:phantom_isolated_motion}
\end{figure}

After reconstruction, motion isolation based on the strategy described in Section \ref{sec:Method} is straightforward. Figure \ref{fig:phantom_isolated_motion} shows the isolated cardiac (top) and respiratory (bottom) motion reconstructions, together with results that were obtained with the ICA-based baseline approach. It can be seen that the proposed method indeed achieves a decomposition into the different types of motion, and that this decomposition yields a much sharper reconstruction of the different motion types compared to the ICA-based approach (see the error images).

\subsection{Cardiac MR images}
We now test our method on cine MR images, comprising two images with a four-chamber view and two images with a short-axis view \footnote{Data from the ISMRM reconstruction challenge 2014 \href{https://challenge.ismrm.org/}{challenge.ismrm.org}.}.

\begin{figure}[t]
	\centering
	\includegraphics[width=0.75 in]{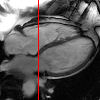} %
	\includegraphics[width=0.75 in]{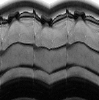}	
	\includegraphics[width=0.75 in]{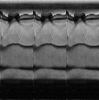}	
	\includegraphics[width=0.75 in]{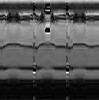}	
	\includegraphics[width=0.75 in]{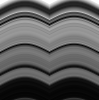}	
	\includegraphics[width=0.75 in]{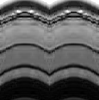}	

	\includegraphics[width=0.75 in]{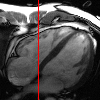} %
	\includegraphics[width=0.75 in]{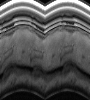}	
	\includegraphics[width=0.75 in]{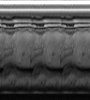}	
	\includegraphics[width=0.75 in]{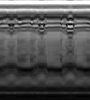}	
	\includegraphics[width=0.75 in]{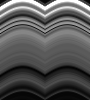}	
	\includegraphics[width=0.75 in]{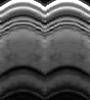}	

	\caption{Four-chamber examples. From left to right: Reference frame with marked slice, compact representation of full dynamics (ground truth), of cardiac dynamics (ground truth and reconstruction) and of respiratory dynamics (ground truth and reconstruction).
\label{fig:four_chamber_results}}	
\end{figure}

\begin{figure}[t]
	\centering
	\includegraphics[width=0.83 in]{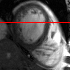}
	\includegraphics[width=0.75 in]{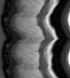}	
	\includegraphics[width=0.75 in]{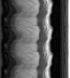}	
	\includegraphics[width=0.75 in]{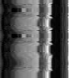}	
	\includegraphics[width=0.75 in]{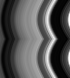}	
	\includegraphics[width=0.75 in]{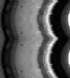}

	\includegraphics[width=0.84 in]{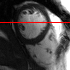} 
	\includegraphics[width=0.75 in]{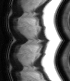}	
	\includegraphics[width=0.75 in]{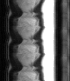}	
	\includegraphics[width=0.75 in]{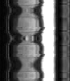}	
	\includegraphics[width=0.75 in]{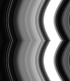}	
	\includegraphics[width=0.75 in]{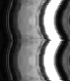}

	\caption{Short-axis examples. From left to right: Reference frame with marked slice, compact representation of full dynamics (ground truth), of cardiac dynamics (ground truth and reconstruction) and of respiratory dynamics (ground truth and reconstruction).\label{fig:short_axis_results}}

\end{figure}

In all experiments considered here, the original videos were obtained via a sum-of-squares reconstruction from fully sampled MR data, and contain a 2D slice of the entire thorax showing the beating heart in one region. For our experiments, we simulated three heartbeats by concatenating the single-heartbeat-videos three times, and simulated two respiratory cycles with vertical (resp. horizontal) shearing motion for the four-chamber (resp. short-axis) view. After obtaining videos showing a slice of the entire thorax with three heartbeats and two respiratory cycles, we cropped the videos to a region of interest around the heart, see the the two left columns of Figures \ref{fig:four_chamber_results} and \ref{fig:short_axis_results}. The final data consists of 99 resp. 90 frames with spatial resolution $100 \times 100$ for the two four-chamber view examples, and of 81 resp. 78 frames with spatial resolution $70 \times 70$ for the two short-axis view examples.

In order to account also for measurement errors in the motion trigger, the motion trigger for respiratory motion provided to our method was a perturbed one with the time time-position where the motion trigger is sampled being perturbed by additive Gaussian noise with a standard deviation of 50\% of the length of one timestep. To account for that, in our method we also allowed deviations from the perturbed trigger that are penalized with an $L^2$ discrepancy.

In all experiments with real data, the generator is a standard deep convolutional neural network (CNN) with 7 layers, ReLU activation functions in the first and last layer, LeakyReLUs in the middle layers and no biases. The latent space $\mathbb{R}^{100}$ is split into 98 static components and 2 dynamic components. %
Network parameters shared by all experiments are given as: Number of channels: $[100,640,320,160,80,40,20,1]$, stride: $[2,2,2,2,2,2,1]$. %
To account for the different image dimensions, the shape of the remaining parameters differs slightly: 
(Square) kernel size: $[4,4,4,4,4,4,3]$ (four-chamber), $[4,4,4,4,4,5,4]$ (short-axis), padding: $[0,2,0,2,2,1,1]$ (four-chamber), $[0,2,2,2,2,1,1]$ (short-axis). In total, the networks have $5.388980 \times 10^{6}$ (four-chamber) and $5.396320 \times 10^{6}$ (short-axis) parameters.
The network is optimized according to Equation \eqref{eq:specific_opt_problem}, where the latent variables are randomly initialized from a uniform distribution on the interval $[0, 1)$, and the network weights are initialized with Pytorch self-initialization. For both experiments, blocks of $[4000,4000,4000,4000,4000]$ epochs with learning rates $[0.01,0.008,0.005,0.003,0.001]$ are used for the Adam optimizer. 
A comparison to the ICA-based approach was also carried out, but since we were not able to obtain reasonable results with the ICA-based approach, the comparison is not included here.

\begin{figure}[t]
	\centering
		\includegraphics[width=0.24\textwidth, trim=1.1cm 0 1.5cm 0, clip]{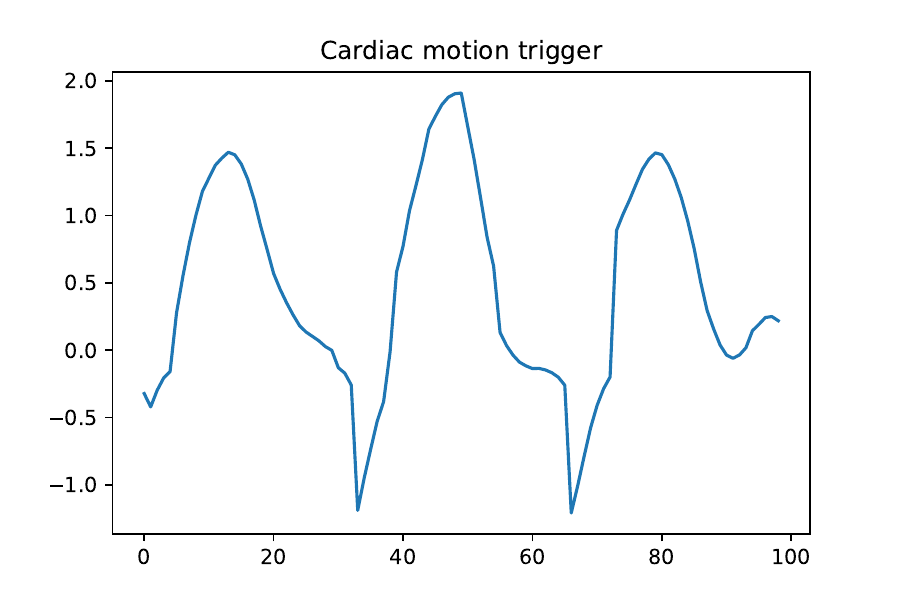}
		\includegraphics[width=0.24\textwidth, trim=1.1cm 0 1.5cm 0, clip]{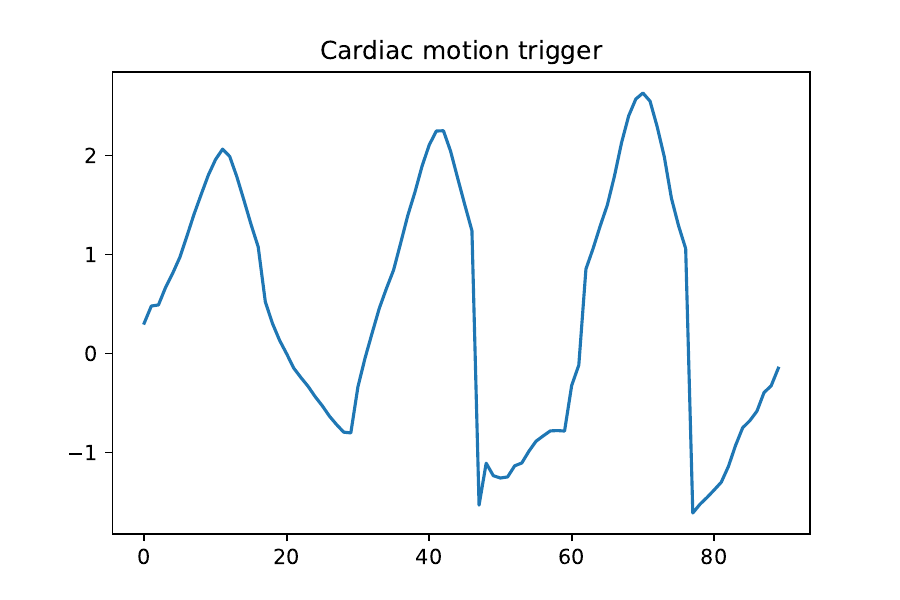}
		\includegraphics[width=0.24\textwidth, trim=1.1cm 0 1.5cm 0, clip]{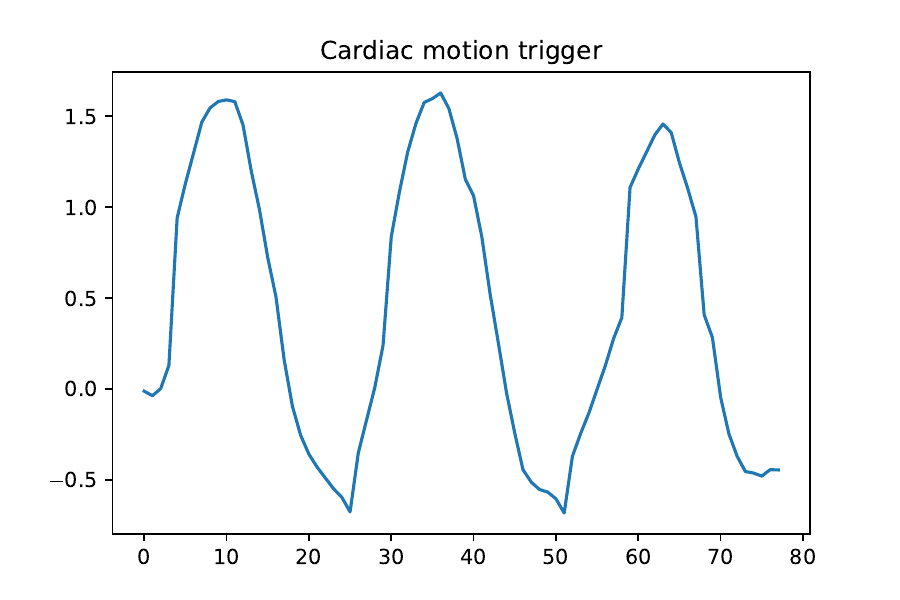}
		\includegraphics[width=0.24\textwidth, trim=1.1cm 0 1.5cm 0, clip]{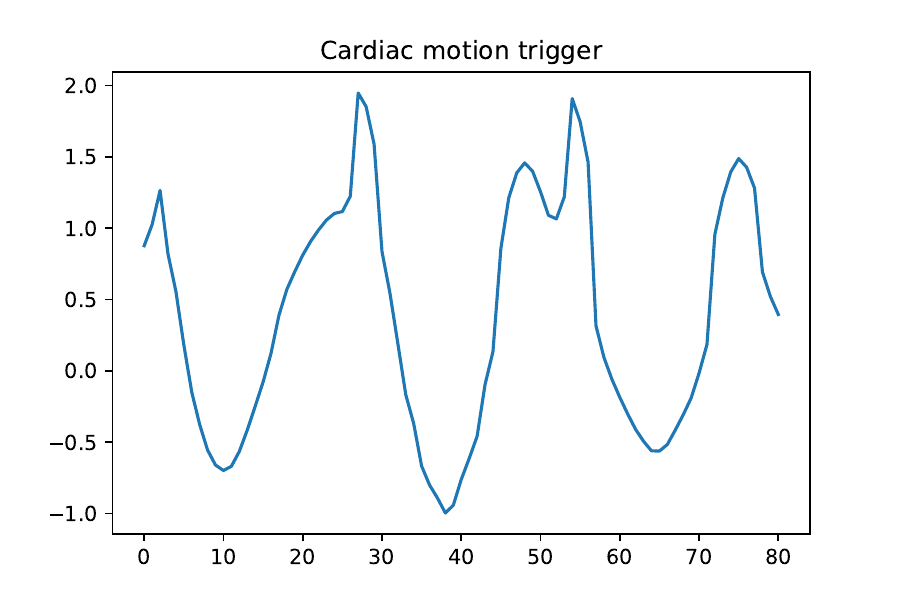}

	\caption{Reconstructed motion triggers for the four-chamber examples (left two images) and the short-axis examples (right two images).\label{fig:perturbed_count}}
\end{figure}

Motion isolation is performed on the MR images as explained in Section \ref{sec:Method}. The learned generator and the reconstructed dynamic latent space variables are used to freeze one type of motion while maintaining the dynamics of the other type of motion. The latent space variables associated to the cardiac motion are shown in Figures \ref{fig:perturbed_count}. Note that these plots indeed seem to reassemble patterns associated to the heart's activity, and were completely unknown before performing the optimization. 

The results of the motion isolation experiments are shown in Figures \ref{fig:four_chamber_results} and \ref{fig:short_axis_results} for the four-chamber view experiments the short-axis view experiments, respectively.
It can be observed that in all cases and for both types of motion, the isolation of motion works well, and the different structures of the motion are clearly visible in the slice-based visualization of the generated images. This is also confirmed by the quantitative values provided in Table \ref{tbl:seed_errors}. 

We should note that some artifacts are visible in the breathing motion isolation. In our experience, those are mostly related to having obtained a sub-optimal solution of the minimization problem (recall that we provide results corresponding to the median of the performance of our method). In cases where a favorable random initialization leads to an improved convergence of the methods (in terms of minimizing the objective functional), these artifacts are reduced. %
Based on this observation, in order to further improve the results, one could in practice monitor the loss during iterations and, in case a suboptimal approximation of the data was achieved, restart the method with a different, random initialization.

\begin{table}[hbtp]
\caption[Parameters]{Error in motion isolation (repeating each experiment for 20 different seeds). MAD denotes the median absolute deviation.}
\centering
{\scriptsize 
\begin{tabular}{ p{6.5cm} p{1.3cm}p{1.3cm}p{1.3cm}p{1.3cm} }
 \toprule
 & Median & MAD & Mean & Std. dev.\\
 \midrule
 Phantom example,  $\mathbf{E}_h^1$ (cardiac) & 1.06e-02 & 2.87e-03 & 1.36e-02 & 8.32e-03 \\
 Phantom example,  $\mathbf{E}_h^2$ (respiratory) & 9.79e-03 & 1.41e-03 & 1.08e-02 & 2.97e-03 \\
 Four-chamber view, $\mathbf{E}_h^1$ (cardiac)&  1.13e-01 & 2.89e-02 & 1.24e-01 & 3.93e-02 \\
 Four-chamber view, $\mathbf{E}_h^2$ (respiratory)& 8.66e-02 & 1.69e-02 & 9.76e-02 & 2.75e-02 \\
 Four-chamber view, example 2, $\mathbf{E}_h^1$ (cardiac)&  9.11e-02 & 6.71e-03 & 1.09e-01 & 5.70e-02 \\
 Four-chamber view, example 2, $\mathbf{E}_h^2$ (resp.)& 7.64e-02 & 1.01e-02 & 8.98e-02 & 4.94e-02 \\
 Short-axis view, $\mathbf{E}_h^1$ (cardiac) &1.42e-01 & 1.88e-02 & 1.46e-01 & 3.38e-02 \\
 Short-axis view, $\mathbf{E}_h^2$ (respiratory) &1.02e-01 & 1.59e-02 & 1.10e-01 & 2.91e-02  \\
 Short-axis view, example 2, $\mathbf{E}_h^1$ (cardiac) &8.26e-02 & 6.87e-03 & 8.38e-02 & 1.22e-02  \\
 Short-axis view, example 2, $\mathbf{E}_h^2$ (respiratory) & 6.91e-02 & 4.91e-03 & 6.94e-02 & 1.03e-02 \\
 \midrule
 Phantom example, ICA-method, $\mathbf{E}_h^1$ (cardiac) &  3.08e-02 & \multicolumn{1}{c}{-}  & \multicolumn{1}{c}{-} & \multicolumn{1}{c}{-}\\%
 Phantom example, ICA-method, $\mathbf{E}_h^2$ (resp.) & 3.87e-02 & \multicolumn{1}{c}{-} & \multicolumn{1}{c}{-} & \multicolumn{1}{c}{-}\\   
 \bottomrule
\end{tabular}
}
\label{tbl:seed_errors}
\end{table}
\normalsize

\paragraph{Conclusions and outlook}
This paper introduces a new method for motion isolation based on the joint optimization of an untrained generator network over both the network parameters and the latent codes. Assuming one-dimensional information on all but one of the motions is known, motion isolation is achieved through latent space disentanglement.
Feasibility of this method was shown for isolating respiratory and cardiac motion in dynamic MR images, but the proposed method is general and can conceptually be used in many applications, e.g., in bio-medical imaging, biology, bio-mechanics or physics.

A limitation of the method, resulting from non-convexity, is its dependence on initializations, which is counteracted here via loss-based restarting strategies. Further, no explicit motion information is made available by our methods. While this might be considered as limitation, it also comes with the advantage that no explicit modeling of the underlying motion types is necessary.
Our work shows a great potential of latent space disentanglement on untrained generators for video data. It opens the door to more advanced disentanglement schemes (e.g., based on modifications of the loss function or additional constraints on the latent space variables). Moreover, we expect the proposed method to be very well suited as image prior for
dynamic inverse problems (e.g., in tomography, super-resolution) where the reconstructed solution displays different kinds of independent motion.

\bibliography{references}
\bibliographystyle{splncs04}

\end{document}